# Answer to Comment by J.B. Pendry on" Left-handed materials do not make a perfect lens"

N. Garcia (1) and M. Nieto-Vesperinas (2)
(1) Laboratorio de Fisica de sistemas Pequeños, Consejo
superior de Investigaciones Cientificas, Serrano 144, Madrid
28006, Spain. E-address:
nicolas.garcia@fsp.csic.es
(2) Corresponding author. E-address: mnieto@icmm.csic.es,
Instituto de Ciencia de Materiales de Madrid, Consejo Superior de
Investigaciones Cientificas, Campus de Cantoblanco, Madrid 28049,
Spain

Ref. 1 does not address absorption for a left-handed material (LHM). The existence of the amplified wave inside such medium (cf. Eqs.(23) and (24) of [1]) is proposed for the ideal lossless case: $\epsilon = \mu = n = -1$. In fact, below Eq.(23) it is stated: "Therefore there is no physical obstacle to perfect reconstruction of the image beyond practical limitations". Absorption is only accounted for in the last example of reconstruction in [1] as a limiting factor to the proposed perfect focusing, but this is illustrated by using a silver slab, not a LHM slab (we discuss this difference below), when it is stated below Eq.(33): "Evidently the imaginary part of the dielectric function will place some practical limitations". Thus, there is no misattribution of [1] when an ideal lossless LHM is addressed in the first part of [2]. In his comment Pendry now suggests absorption to keep his perfect lens proposal of [1] free from the divergencies that, as shown in [2], invalidate the aforementioned amplified wave model. However, contrary to what it is stated in the last paragraph of Pendry's comment, [2] showed that absorption of course avoids those divergencies, but leads to an evanescent wave and thus prevents perfect lens focusing as well.

The problem with the new proposal in Pendry's comment is the following: In order that a plasmon in the second surface of the slab be the agent of the amplified wave as stated in [1] and in the present comment, this slab must be of metal in the electrostatic limit. Which, as shown in Eq.(27) of [1], implies that $k_0 << \sqrt{k_x^2 + k_y^2}$, ($k_0 = \omega/c$). This excludes low values of $k_x$. Therefore, the summation in $k_x$ performed in Eq.(34) of [1] to reach the simulated reconstruction of its Fig.2(c) is incorrect because although not explicitly stated in that equation, the only way to get such a reconstruction is to perform that summation by including *all* $k_x$ components, namely, both those that are in the propagating region : $-k_0 \leq k_x \leq k_0$, and those in the evanescent region that are closer to $k_0$; so that the premise condition: $\omega << c\sqrt{k_x^2 + k_y^2}$ is violated. Notice also that, out from the electrostatic limit, in the silver slab proposed in [1] the $k_z$ wavevector component is not $k_z = i\sqrt{k_x^2 + k_y^2 - k_0^2}$ as for a LHM slab, but: $k_z = i\sqrt{k_x^2 + k_y^2 + k_0^2}$, and only under the above quoted electrostatic limit condition, both expressions of $k_z$ coincide, this occurs approximately at: $k_x = 5k_0$. Now, if there is absorption , the image spectrum, (i.e., the terms in the sum of Eq.(34) of [1]) for those values of $k_x$ at which the condition $k_0 << \sqrt{k_x^2 + k_y^2}$ is satisfied, (namely, for $k_x \geq 5k_0$), decays for $\epsilon = -1 + 0.4i$ almost by a factor $10^{-2}$ of its value at $k_x = k_0$. Thus absorption gives practically no signal in the image plane. Notice as well that this decay of the image spectrum in the electrostatic limit is also a consequence of the fact that then the condition $\delta exp(|k_z|d) >> 1$, established in [2] (cf. [2] just before Eq.(6)), for an absorbing LHM, holds for $k_z = i\sqrt{k_x^2 + k_y^2}$ and $k_0 << \sqrt{k_x^2 + k_y^2}$, and thus the evanescent components inside the medium are decaying. In other words, the condition (3) proposed in Pendry's comment for a silver layer is not satisfied in the electrostatic limit.

We remark that this theory applies to an ideal effective continuous medium of LHM, assuming it can ever be built within the decay length of an evanescent wave. Notice, however, that the photonic crystal structure of so far developed metamaterials possesses a lattice period that limits both the evanescent wave $k$-vector and the bulk space at which the evanescent component "sees" such an effective medium, then making meaningless the continuous model of [1] and their divergencies. Nevertheless, it remains to study whether ultrathin absorbing layers of metal, or of LHM, can restore evanescent components at any $k_z$ not in the electrostatic limit when much smaller distances than those proposed in [1] are used. Nevetheless, one should place a detector at separations smaller than 10 nm from the object. This would involve to interpose this layer within such an interspace, and separate it from both the object and the detector less than 3 nm. A procedure that any near field optical microscopy experimentalist can question as an unnecessary complication to yield no substantial gain upon the free space propagation evanescent wave filter in such near zone.

[1] J.B. Pendry, Phys. Rev. Lett. **85**, 3966 (2000).
[2] N. Garcia and M. Nieto-Vesperinas, Phys. Rev. Lett. **88**, 207403 (2002)